\documentclass[11pt,letterpaper,oneside,pdftex]{article}
\pdfoutput=1
\newcommand{\version}{June 21, 2021}
\usepackage[utf8]{inputenc}
\usepackage[T1]{fontenc}
\usepackage{amsmath}
\usepackage{amsfonts}
\usepackage{amssymb}
\usepackage{mathtools}
\usepackage{graphicx}
\usepackage[width=16.70cm, height=22.50cm]{geometry}
\usepackage{fouriernc}

\usepackage[margin=1cm,font=small]{caption}
\usepackage[numbers,square,comma,sort&compress]{natbib}
\usepackage[pdftex,hyperref,svgnames]{xcolor}
\usepackage[pdftex,bookmarksnumbered=true,breaklinks=true]{hyperref}
\makeatletter
\AtBeginDocument{
	\hypersetup{
		pdftitle = {\@title},
		pdfauthor = {\@author}
	}
}
\makeatother
\makeatletter

\@addtoreset{equation}{section}
\makeatother

\renewcommand{\d}{\delta}

  \newcommand{\cO}{\mathcal{O}}

\newcommand{\pa}{\partial}
\newcommand{\inv}[1]{\frac{1}{#1}}

\newcommand{\nn}{\nonumber}
\newcommand{\eqnref}[1]{Eq. \eqref{#1}}

\newcommand{\kb}{k_{\txt{B}}}

\newcommand{\ct}{c_{\textrm{t}}}

\newcommand{\bt}{\beta_{\textrm{t}}}

\newcommand{\txt}[1]{\textrm{#1}}

\title{\texorpdfstring{\begin{flushright}
			{\small LA-UR-21-23521}
		\end{flushright}\vspace{2em}}{}%
	On the temperature and density dependence of dislocation drag from phonon wind}
\author{Daniel N. Blaschke, Leonid Burakovsky, and Dean L. Preston}
\date{\version}

\graphicspath{{./}{./figures/}}
\begin{document}

\maketitle

\thispagestyle{empty}
\begin{center}
	\vspace{-0.3cm}
	Los Alamos National Laboratory\\Los Alamos, NM, 87545, USA
	\\[0.5cm]
	{\ttfamily{E-mail: dblaschke@lanl.gov, burakov@lanl.gov, dean@lanl.gov}}
\end{center}


 \begin{abstract}
 At extreme strain rates, where fast moving dislocations govern plastic deformation, anharmonic phonon scattering imparts a drag force on the dislocations.
 In this paper, we present calculations of the dislocation drag coefficients of aluminum and copper as functions of temperature and density.
 We discuss the sensitivity of the drag coefficients to changes in the third-order elastic constants with temperature and density.
 \end{abstract}

 \vspace{1cm}
\tableofcontents
 \newpage

\section{Introduction and background}

Plastic deformation at extreme strain rates, $\dot{\varepsilon}>10^5$s$^{-1}$, is governed by dislocation motion that is subject to the dissipative effect of scattering phonons (termed `phonon wind') \cite{Blaschke:2019a,Blaschke:2021impact,Lloyd:2014JMPS,Luscher:2016,Austin:2018}, which in this regime is poorly understood. 
In this work we aim at shedding more light on the temperature and density dependence of this important effect which together with the applied stress determines the speed at which dislocations move through a crystal and consequently the plastic deformation rate that can be achieved at a given dislocation density.

According to Orowan's well known relation \cite{Orowan:1934a,Orowan:1934b,Orowan:1934c}
\begin{align}
	\dot\varepsilon = b \varrho_m v(\sigma)
	\,,
\end{align}
the plastic strain rate $\dot\varepsilon$ is proportional to the density of mobile dislocations $\varrho_m$ and their average velocity as a function of stress.
We are presently interested in temperatures on the order of, or higher than, room temperature.
In the thermally activated (low stress) regime, this average dislocation velocity is dominated by the time dislocations need to overcome various obstacles (other dislocations, defects, grain boundaries, etc.), but with increasing stress these `wait times' become shorter and the gliding velocity that dislocations can achieve between obstacles becomes more important.
As dislocations glide, phonons scatter off them, thereby impeding their motion.
This dissipative effect, known as dislocation drag from `phonon wind', limits dislocation velocities at high stresses and temperatures.
By definition, the dislocation drag coefficient $B = F/v$  is the drag force $F$ per unit length on the dislocation (which is determined from the local stress) divided by its velocity $v$.
The analogy of `drag' is due (historically) to the fact that there is a `viscous' regime where $B$ is almost constant, with typical dislocation velocities in this regime being a few percent of the lowest sound speed in the material \cite{Nadgornyi:1988}.
As the stress increases and dislocation velocities approach the various sound speeds of the crystal, $B$ becomes a highly non-linear function of $v$; see Ref. \cite{Gurrutxaga:2020} for a recent review on dislocation dynamics.

Material strength is the key component of numerous engineering applications and hydrocode simulations.
It has been widely assumed that yield strength, $Y$ (the highest elastic stress at a given strain rate, beyond which plastic flow starts), scales with shear modulus, $G$, as $Y/Y_0 = G/G_0$, where subscript 0 indicates some reference state, usually the one at ambient conditions.
This scaling can be derived theoretically, in the linear elasticity framework, assuming that plastic flow is mainly governed by dislocation motion, and has been confirmed experimentally for several substances at low to moderate compression.
Very recently, however, a sophisticated experimental cross-platform examination of lead strength to 400 GPa \cite{Krygier:2019}, tantalum strength to 350 GPa \cite{Brown:2020}, and copper strength to 50 GPa \cite{Ravindran:2021} revealed that for all three, $Y/Y_0 \approx (G/G_0)^h$ , $h\neq1$; specifically, $h \sim 1.5$ for lead and $\sim 2$ for both tantalum and copper.
Thus, the value of $h$ is material dependent and may have some structure across the periodic table.
To elucidate the source of the additional hardening that leads to this non-linear $Y$-$G$ scaling in the case of tantalum, several alternative mechanisms 
were 
proposed: strain hardening, non-Schmidt stress components, homogeneous dislocation nucleation, twinning, phonon drag, and a phase transformation.
There are compelling reasons to believe that phonon drag may be one of 
the leading mechanisms, or even the sole mechanism responsible for additional 
hardening. 
As mentioned above, at sufficiently high strain rates dislocation velocities are governed by the interaction of dislocations with lattice phonons.
Many models of the phonon drag suggest that $B$ increases with temperature and varies with density via a term related to the longitudinal wave speed.
It is possible that the strength in the viscous drag regime scales non-linearly with the shear modulus.
This is so because the phonon drag coefficients are functions of both second- and third-order elastic constants and, in contrast to the former, the latter do not generally exhibit a linear dependence on pressure and/or temperature.
Thus, we will consider phonon drag as 
a plausible source of additional hardening leading to the non-linearity of $Y$-$G$ scaling, which gives more motivation to our present study.
We note, however, that in order to elucidate the explicit role of phonon drag in the cause of the non-linearity of the $Y$-$G$ scaling a more specific study may be required, 
with more detailed comparison to experiment, etc. 

In a series of papers \cite{Blaschke:BpaperRpt,Blaschke:2018anis,Blaschke:2019fits,Blaschke:2019Bpap} that generalize the earlier work of Alshits et al. \cite{Alshits:1992}, a first-principles theory of dislocation drag from phonon wind, $B(\vartheta,v)$, as a function of dislocation character angle $\vartheta$ and velocity $v$, was developed for fast moving dislocations in arbitrary anisotropic crystals.
In this theory only the phonon spectrum was approximated as isotropic; the dislocation field and all elastic constants are anisotropic, and the actual slip systems are taken into account.

A numerical implementation of this theory is published in the open source code PyDislocDyn \cite{pydislocdyn}.
The crucial ingredients of this dislocation drag theory are the elastic constants of both second-order (SOEC) and third-order (TOEC), as these determine the strength of the dislocation-phonon interaction.
Elastic constants, like most material parameters, are functions of temperature and density (or pressure).
While material density as a function of pressure and temperature is determined by the equation of state \cite{Wallace:2002}, less is known about some of the other material parameters we need to determine $B(v,T,P)$.
Models exist describing the temperature and density dependencies of the polycrystal average bulk and shear moduli; see e.g. \cite{Preston:1992,Burakovsky:2003,Garai:2007,Brown:2020} and references therein, but not for general anisotropic elastic constants.
Furthermore, experimental data --- especially for third-order elastic constants at temperatures higher than room temperature --- are scarce, which limits the accuracy with which the temperature and density dependencies of the drag coefficients can be determined.
A rough first-order approximation of this dependence, where the TOECs were held constant, was discussed in Ref. \cite{Blaschke:2019a} for the isotropic limit and in Ref. \cite{Blaschke:2021impact} for anisotropic crystals.
Here we aim to fill this gap by calculating both SOEC and TOEC at different temperatures and densities for the examples of copper and aluminum using computer simulations via the well-known Vienna Ab initio Simulation Package \cite{Kresse:1993,Kresse:1994a,Kresse:1996a,Kresse:1996b} (VASP, see
\url{www.vasp.at}),
the details of which we describe in the next section.

\section{Determining elastic constants with computer simulations}
\label{sec:VASP}

The second- and third-order elastic constants (SOEC and TOEC) for a number of temperatures and pressures may be calculated along the lines
of Refs.~\cite{Soderlind:1993,Wang:2014,Gu:2019} and \cite{Lopuszynski:2007,Zhao:2007,Wang:2009,Wen:2017}.
The deformations employed in Refs.~\cite{Soderlind:1993,Wang:2014,Gu:2019} are volume preserving, hence they cannot give us all elastic constants as we now explain in detail.

A deformation of a material is described by the gradient $u_{i,j}=\pa_j u_i$ of the continuous displacement field $u_i$.
The finite Lagrangian strain tensor is then given by
\begin{align}
	\eta_{ij}=\inv2\left(u_{i,j}+u_{j,i}+u_{k,i}u_{k,j}\right)
	\,. \label{eq:strain}
\end{align}
Assuming small strains, we may expand the Gibbs free energy $F$ of the deformed material in powers of Lagrangian strain, thereby defining the elastic constants at a given order via the expansion coefficients:
\begin{align}
	\Delta F &= \frac{V}{2!}C_{ijkl}\eta_{ij}\eta_{kl} + \frac{V}{3!}C_{ijklmn}\eta_{ij}\eta_{kl}\eta_{mn} + \ldots
	\label{eq:freeE}
\end{align}
where $V$ is the volume of the unstrained lattice~\cite{Wen:2017} and summation over repeated indices is implied.
Note that the linear term is eliminated by the equilibrium condition at zero strain.
Here, we only consider second- and third-order elastic constants, $C_{ijkl}$ (SOEC) and $C_{ijklmn}$ (TOEC).
Since the strain tensor is symmetric by construction, the elastic constants are symmetric in each index pair and also under exchange of any index pair (Voigt symmetry).
In general, one has up to 21 independent SOECs and up to 56 independent TOECs~\cite{Wallace:1970}.

Presently, we consider only cubic I symmetry in which case the SOEC consist of only 3 independent components and the TOEC consist of only 6 independent components.
The strategy is to calculate the energy changes due to different types and magnitudes of strains via computer simulations and subsequently fit second- and third-order polynomials to the data points.
The coefficients of the polynomials give us the SOEC and TOEC.
For example, in order to get the three independent SOEC we need three different strain configurations and for each one the magnitude of the deformation is varied.

Let us look at the possible strain configurations in more detail.
In the undeformed lattice, the deformation matrix $\alpha_{ij}=\d_{ij}$, and the strain tensor vanishes due to the relation $u_{i,j}=\alpha_{ij}-\d_{ij}$ between the deformation matrix and the displacement gradient \cite{Wallace:1985,Blaschke:2017tpfe}.
A small deviation may be expressed as $\alpha_{ij}=\d_{ij}+x_{ij}$, and restricting ourselves to symmetric small $x_{ij}$, the latter have six independent components which in Voigt notation read $x_1,\ldots,x_6$.
We remind the reader that Voigt notation maps index pairs to single digits: $(11, 22, 33, 32/23, 31/13, 21/12) \rightarrow (1, 2, 3, 4, 5, 6)$.
The six independent strain components $\eta_1,\ldots,\eta_6$ in Voigt notation may be expressed in terms of second-order polynomials in the $x_i$; see \eqref{eq:strain}.
Substituting these expressions into \eqref{eq:freeE} and truncating the expansion at third order (or even second order if we seek only SOEC), gives us the required polynomials.
Careful choices of $x_i$ lead to polynomials depending only on a subset of elastic constants.

We begin in the following subsection by examining volume preserving deformations where $\det \alpha_{ij}=1$.
As usual, the names of the independent components of the elastic constant tensors are inspired by Voigt notation.

\subsection{Volume preserving deformations}
\begin{enumerate}
	\item Choosing $x_1=x_2=x_4=x_5=0$, $x_3=\frac{1}{(1-y^2)}-1$, and $x_6=y$ yields $\det \alpha_{ij}=1$ and
	\begin{align}
		\Delta F(y)=2C_{44}y^2 + \cO(y^4)
		\,. \label{eq:c44sym}
	\end{align}

	\item Choosing $x_1=y=-x_2$, $x_3=\frac{1}{(1-y^2)}-1$, and $x_4=x_5=x_6=0$ yields $\det\alpha_{ij}=1$ and
	\begin{align}
		\Delta F(y)=(C_{11} - C_{12})y^2  + \cO(y^4)
		\,. \label{eq:cprsym}
	\end{align}
	
	\item Choosing $x_1=x_2=y$, $x_3=\frac{1}{(1+y)^2}-1$, and $x_4=x_5=x_6=0$ yields $\det\alpha_{ij}=1$ and
	\begin{align}
		\Delta F(y)=3 (C_{11} - C_{12})y^2 - 9(C_{11}-C_{12})y^3 - (C_{111} - 3C_{112} + 2C_{123})y^3 + \cO(y^4)
		\,. \label{eq:cprime}
	\end{align}
	
	\item Choosing $x_1=x_2=x_3=0$ and $x_4=x_5=x_6=y$ yields $\det\alpha_{ij}=1$ and
	\begin{align}
		\Delta F(y)=6C_{44}y^2 + 6C_{44}y^3 + 8C_{456}y^3 + \cO(y^4)
		\,. \label{eq:c44}
	\end{align}
	
	\item Choosing $x_2=y=-x_3$, $x_4=x_5=y$, and $x_1=x_6=0$ yields $\det\alpha_{ij}=1$ and
	\begin{align}
		\Delta F(y)=(C_{11} - C_{12} + 4C_{44})y^2 -\frac12 (C_{11} - C_{12} + 4C_{44})y^3 + 2(C_{144} - C_{166})y^3 + \cO(y^4)
		\,. \label{eq:TOEC46}
	\end{align}
	
\end{enumerate}
One can show that any other volume preserving deformation leads to polynomials depending on the same two SOEC and same three TOEC, namely
$(C_{11}-C_{12})/2$, $C_{44}$, $C_{456}$, $(C_{144}-C_{166})$, and $(C_{111}-3C_{112}+2C_{123})$.

\subsection{Volume non-preserving deformations}
The remaining SOEC, the bulk modulus $K=(C_{11}+2C_{12})/3$, and one additional TOEC, $(C_{111} + 6C_{112} + 2C_{123})$, can be calculated from a rescaling deformation of the metal lattice.
Choosing $x_1=x_2=x_3=y$ and $x_4=x_5=x_6=0$ corresponds to a rescaling of the lattice constant by $a\to(1+y)a$ and yields $\det\alpha_{ij}=(1+y)^3$ and
\begin{align}
	\Delta F(y)= (1.5C_{11} + 3C_{12})y^2 + (1.5C_{11} + 3C_{12})y^3 + (0.5C_{111} + 3C_{112} + C_{123})y^3 + \cO(y^4)
	\,. \label{eq:rescale}
\end{align}
In the case of the bulk modulus, better accuracy is achieved by calculating the change in pressure as a function of material density $\rho(y)$ and taking the first derivative of a fitting polynomial at $y=0$.

In order to calculate the remaining two TOEC we must resort to deformations that do not preserve the volume.
Our choices are 
\begin{enumerate}
	\item $x_1=y$ and all other $x_i=0$:
	\begin{align}
		\Delta F(y)= 0.5C_{11}y^2 + 0.5C_{11}y^3 + \frac16C_{111}y^3 + \cO(y^4)
		\, , \label{eq:TOEC1}
	\end{align}
	
	\item $x_1=x_2=y=x_4$ and $x_3=x_5=x_6=0$:
	\begin{align}
		\Delta F(y)&= (C_{11} + C_{12} + 2C_{44})y^2 + (1.5C_{11} + 2.5C_{12} + 2C_{44})y^3\nn\\
		&\quad + (\tfrac13C_{111} + C_{112} + 2C_{144} + 2C_{166})y^3 + \cO(y^4)
		\,. \label{eq:TOEC6}
	\end{align}
	
\end{enumerate}


\begin{figure*}[ht]
	\centering
	\includegraphics[width=0.503\textwidth]{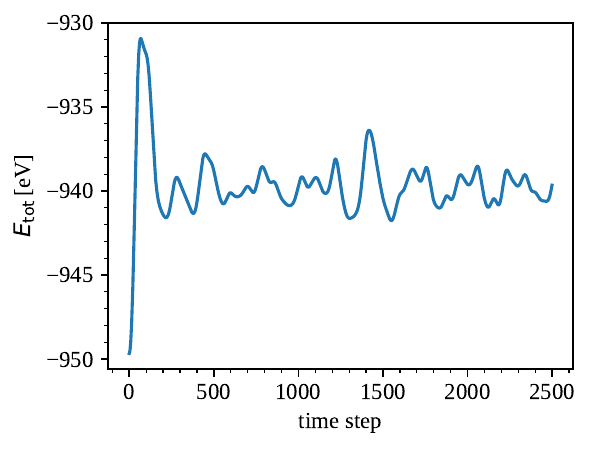}%
	\includegraphics[width=0.497\textwidth]{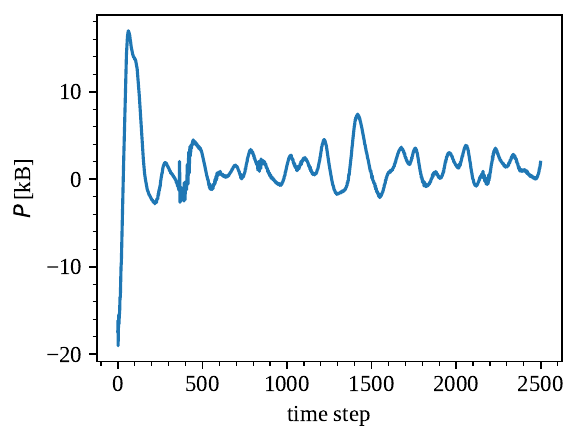}
	\caption{Energy (left) and pressure (right) as a function of simulation time step (in fs) calculated with VASP (Vienna Ab initio Simulation Package) for copper at 300K and with lattice constant $a=3.647$\r{A} for an undeformed fcc lattice containing 256 atoms.
		Note that the experimental value for the lattice constant at ambient pressure is $a=3.6146$\r{A}, the difference being due to the assumed potential used for the simulation.}
	\label{fig:EPref_Cu_300amb}
\end{figure*}

\begin{figure*}[ht]
	\centering
	\includegraphics[width=0.505\textwidth]{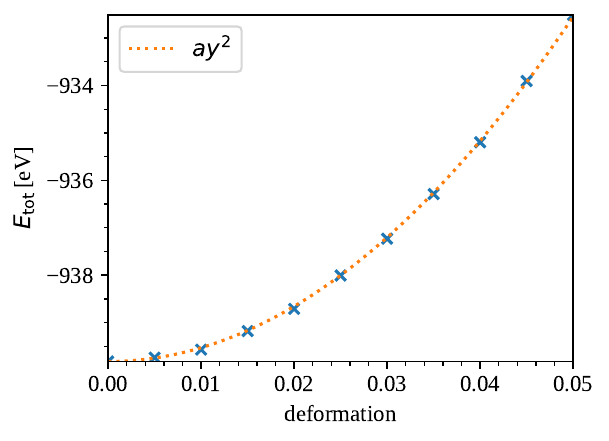}%
	\includegraphics[width=0.495\textwidth]{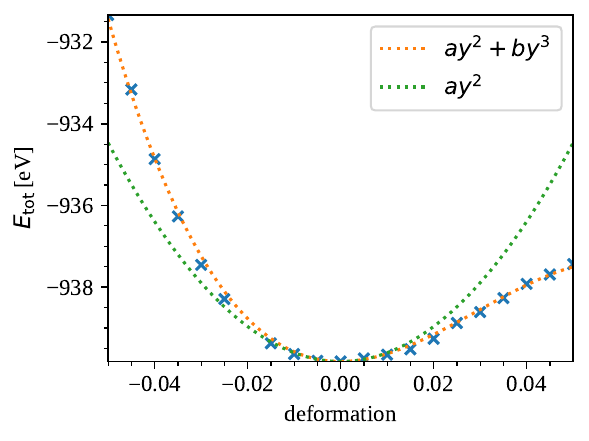}
	\caption{Energy as a function of deformation $y$ for deformation types \eqref{eq:c44sym} (left) and \eqref{eq:cprime} (right), and including the fitting functions for SOEC and TOEC.
		Since \eqref{eq:c44sym} is symmetric in $y$ up to third order (i.e. it does not depend on $y^3$, only positive points are required, and we can only get one SOEC, $C_{44}$ in this case.
		Eq. \eqref{eq:cprime} is very asymmetric, allowing us to extract both a SOEC (proportional to $a$) and a TOEC (proportional to $b$).}
	\label{fig:c44sym+cprime}
\end{figure*}

\begin{figure*}[ht]
	\centering
	\includegraphics[width=0.51\textwidth]{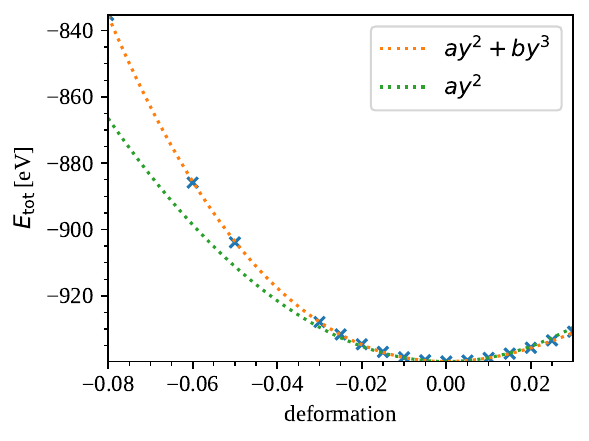}%
	\includegraphics[width=0.49\textwidth]{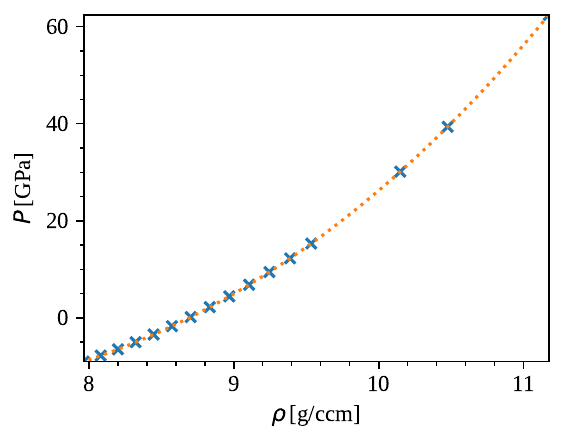}
	\caption{Energy (left) as a function of deformation of the rescaling type \eqref{eq:rescale}, and pressure (right) as a function of density.
		The plots include the fitting functions for SOEC, TOEC (left), and an ``equation of state'' polynomial (right) whose derivative at ambient density ($\rho=8.70$g/cm$^3$ in the simulation) yields the bulk modulus with slightly better accuracy than from the energy plot on the left.
		The latter, however, is required to calculate the TOEC $(C_{111} + 6C_{112} + 2C_{123})$ which is related to coefficient $b$ in the figure legend.}
	\label{fig:bulk}
\end{figure*}

\begin{figure*}[!h!t]
	\centering
	\includegraphics[width=0.5\textwidth]{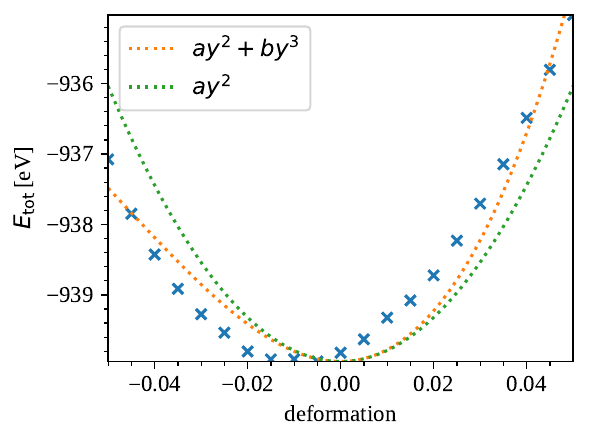}%
	\includegraphics[width=0.5\textwidth]{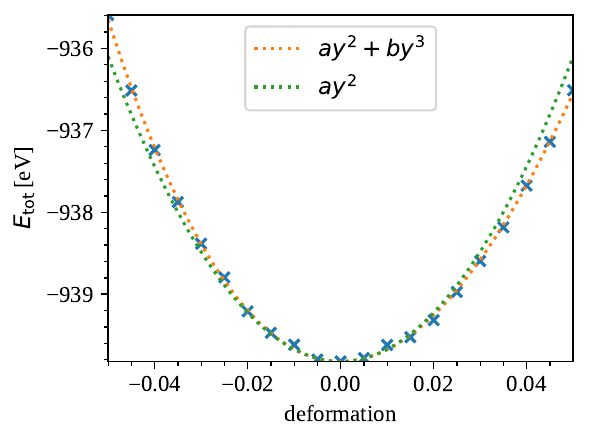}%
	\caption{Energy as a function of deformation $y$ for deformation 
		type \eqref{eq:TOEC1} 
		for copper at room temperature and ambient pressure, and including the fitting functions for SOEC and TOEC.
		Note (left plot) that even though the data were corrected by $P\mathrm{d}V$, the lowest energy value is still displaced from zero deformation.
		Only after including the (volume-dependent) vibrational Helmholtz free energy contribution to $E_\txt{tot}$ (discussed in the text) do we get 
		a curve whose minimum is located at zero deformation
		which enables us to compute the TOECs, albeit with significantly greater uncertainty because these constants are very sensitive to those corrections.}
	\label{fig:TOEC1-TOEC6}
\end{figure*}

\subsection[Using VASP to calculate energy and pressure as a function of \texorpdfstring{$y$}{y}]{Using VASP to calculate energy and pressure as a function of \texorpdfstring{$\mathbf{y}$}{y}}

For our VASP simulations of the elastic constants of copper and aluminum, we used the generalized gradient approximation (GGA) with the Perdew-Burke-Ernzerhof (PBE) exchange-correlation functional \cite{Perdew:1996}.
We modeled Cu and Al using the following core-valence representations:\\
{[$^{18}$Ar]3d$^{10}$4s$^1$} for Cu and [$^{10}$Ne]3s$^2$3p$^1$ for Al, i.e., we assigned, respectively, the 11 and 3 outermost electrons to the valence.
The valence electrons were represented with a plane-wave basis set with cutoff energies of, respectively, 400 and 300 eV, while the core electrons were represented by projector augmented-wave (PAW) pseudopotentials \cite{Bloechl:1994,Kresse:1999}.
We used a $4\times 4\times 4$ (256-atom) supercell in each case, with a single $\Gamma $-point and periodic boundary conditions.
Full energy convergence (to $\lesssim 1$ meV/atom) was checked for each run.
The finite-$T$ simulations (where ions are allowed to move during the simulation) typically require a few thousand time steps (of 1 fs) to achieve full energy convergence and to produce sufficiently long output for the extraction of reliable averages for the values of both energy and pressure.
Fig. \ref{fig:EPref_Cu_300amb} shows the first 2500 time steps in one of our simulations.

For each of the above deformations, Eqns. \eqref{eq:c44sym}--\eqref{eq:TOEC6}, we have calculated $y$ from $\pm0.005$ to $\pm0.05$ in steps of $0.005$ (except for the symmetric polynomials where negative $y$ were omitted) for both copper and aluminum at several temperatures $T\ge300$K and pressures.
For each one of these deformations, a separate simulation was carried out.
A simulation cell of 256 atoms in a fcc structure was used to calculate energies and pressures.
In order to obtain sufficiently accurate equilibrium averages the simulations were carried out to 2500 time steps.
The first 500 steps were discarded in postprocessing and the remaining 2000 were averaged over to obtain one energy resp. pressure.
We found that the calculated pressure at a given initial value of the lattice constant differs from the experimental value due to approximations in the potential.
In other words, the room temperature, low pressure value for the lattice constant in the simulation differs slightly from the experimental value.
In calculating elastic constants, this needs to be corrected for either by accounting for the different pressure, or by changing the lattice constant.

Figure~\ref{fig:EPref_Cu_300amb} is an example of the time dependence of the computed energy and pressure at a given lattice constant, temperature, and in this sample plot without any deformation.
Similar data were also calculated using VASP for all of the deformations described above.
In each case, both energy and pressure were determined by averaging over all time steps after relaxation had occurred.
The first 200-300 time steps clearly needed to be removed, and in fact we discarded the first 500 in order to be on the safe side.
From Figure~\ref{fig:EPref_Cu_300amb} we find a pressure of 0.13 GPa.
This is of course higher than atmospheric pressure, but the effect on the computed elastic constants is still smaller than the inherent uncertainty in the VASP calculations, namely $\sim$0.1 GPa for SOECs and even less for TOECs.  
Collecting the average energies and densities from numerous simulations allows us to plot energy or pressure versus deformation of a given type.
Figures~\ref{fig:c44sym+cprime} and \ref{fig:bulk} are representative examples of such plots.
Third-order fitting polynomials then yield the elastic constants as outlined above in eqns \eqref{eq:c44sym}--\eqref{eq:TOEC6}.
Note that the energies first need to be converted to GPa using the conversion factor
\begin{align}
	X = N a^3/160.2177 n
\end{align}
where $N=256$ is the total number of atoms in the simulation, $a$ is the lattice constant, $n$ is the number of atoms per unit cell ($n=4$ for an fcc lattice), and the numerical factor converts eV/\r{A} to GPa.

\begin{table*}[h!t!b]
{\renewcommand{\arraystretch}{1.2}
\small
\centering
\begin{tabular}{l|c|c|c|c}
	$T=300$K & Cu (calculated) & Cu (experiment) & Al (calculated) & Al (experiment)\\\hline
	$\rho$ [g/cm$^3$] & 8.701 & 8.960 & 2.668 & 2.700 \\
	$a$ [\r{A}] & 3.647 & 3.615 & 4.065 & 4.050 \\
	$P$ [GPa] & 0.13 & 0.0001 & 0.04 & 0.0001 \\\hline
	$(C_{11}-C_{12})/2$ [GPa] & 18.41 & 23.55 & 23.79 & 23.17 \\
	$C_{44}$ [GPa] & 75.04 & 75.70 & 34.46 & 28.34 \\
	$K=(C_{11}+2C_{12})/3$ [GPa] & 130.68 & 136.90 & 69.64 & 75.86 \\\hline
	$C_{111}-3C_{112}+2C_{123}$ [GPa] & 921 & 1071 & 1433 & $-59$ \\
	$C_{456}$ [GPa] & 47 & $-95$ & $-18$ & $-30$ \\
	$C_{144}-C_{166}$ [GPa] & 542 & 777 & $-8$ & 317 \\
	$C_{111} + 6C_{112} + 2C_{123}$ [GPa] & $-7416$ & $-6255$ & $-3439$ & $-2894$ \\
	$C_{111}$ [GPa] & $-1574$ & $-1271$ & 925 & $-1076$ \\
	$\tfrac13C_{111} + C_{112} + 2(C_{144} + C_{166})$ [GPa] & $-2840$ & $-2804$ & $-1607$ & $-1400$
\end{tabular}
\caption{Comparison of calculated and experimental SOECs and TOECs at room temperature.
The experimental values were taken from the CRC handbook, Ref.~\cite{CRCHandbook}.
Note that the experimental TOECs have large uncertainties.
The first two lines compare the lattice constants used in the simulation and the resulting calculated pressure to the experimental lattice constants at ambient pressure.}
\label{tab:results-300K}
}
\end{table*}

\begin{table*}[h!t!b]
{\renewcommand{\arraystretch}{1.2}
\small
\centering
\begin{tabular}{l|c|c|c|c}
	$T=300$K & Cu & Cu & Cu & Cu \\\hline
	$\rho$ [g/cm$^3$] & 8.788 & 8.875 & 9.331 & 10.024 \\
	$a$ [\r{A}] & 3.6350 & 3.623 & 3.5630 & 3.4790 \\
	$P$ [GPa] & 1.47 & 2.88 & 11.08 & 26.87 \\\hline
	$(C_{11}-C_{12})/2$ [GPa] & 18.68 & 19.54 & 23.47 & 29.67 \\
	$C_{44}$ [GPa] & 77.80 & 80.44 & 98.79 & 129.99 \\
	$K=(C_{11}+2C_{12})/3$ [GPa] & 137.70 & 144.87 & 185.52  & 257.31 \\\hline
	$C_{111}-3C_{112}+2C_{123}$ [GPa] & 919 & 975 & 1158 & 1436 \\
	$C_{456}$ [GPa] & 48 & 49 & 51 & 51 \\
	$C_{144}-C_{166}$ [GPa] & 564 & 605 & 718 & 936 \\
	$C_{111} + 6C_{112} + 2C_{123}$ [GPa] & $-8045$ & $-7641$ & $-10015 $ & $-12739$ \\
	$C_{111}$ [GPa] & $-1698$ & $-1862$ & $-2700$ & $-3417$ \\
	$\tfrac13C_{111} + C_{112} + 2(C_{144} + C_{166})$ [GPa] & $-2952$ & $-3074$ & $-3829$ & $-4951$
\end{tabular}
\caption{Calculated SOECs and TOECs for Cu at different pressures at room temperature.}
\label{tab:results-Cu-300K_highP}
}
\end{table*}

\begin{table}[h!t!b]
{\renewcommand{\arraystretch}{1.2}
\small
\centering
\begin{tabular}{l|c|c|c}
	$T=300$K & Al & Al & Al\\\hline
	$\rho$ [g/cm$^3$] & 2.965 & 3.660 & 5.211 \\
	$a$ [\r{A}] & 3.9247 & 3.6585 & 3.2520 \\
	$P$ [GPa] & 9.34 & 45.05 & 200.96 \\\hline
	$(C_{11}-C_{12})/2$ [GPa] & 41.18 & 90.35 & 156.37 \\
	$C_{44}$ [GPa] & 54.61 & 139.73 & 420.91 \\
	$K=(C_{11}+2C_{12})/3$ [GPa] & 109.54 & 242.58 & 707.95 \\\hline
	$C_{111}-3C_{112}+2C_{123}$ [GPa] & 2154 & 3417 & 7753 \\
	$C_{456}$ [GPa] & $-60$ & $-159$ & $-84$\\
	$C_{144}-C_{166}$ [GPa] & 94 & 452 & 2220 \\
	$C_{111} + 6C_{112} + 2C_{123}$ [GPa] & $-4911$ & $-8788$ & $-21478$ \\
	$C_{111}$ [GPa] & $-602$ & $-3081$ & $-3334$ \\
	$\tfrac13C_{111} + C_{112} + 2(C_{144} + C_{166})$ [GPa] & $-2761$ & $-5580$ & $-12212$
\end{tabular}
\caption{Calculated SOECs and TOECs for Al at different pressures at room temperature.}
\label{tab:results-Al-300K_highP}
}
\end{table}

\begin{table}[h!t!b]
{\renewcommand{\arraystretch}{1.2}
\small
\centering
\begin{tabular}{l|c|c|c}
	$T=600$K & Cu & Cu & Cu\\\hline
	$\rho$ [g/cm$^3$] & 8.701 & 9.331 & 10.024 \\
	$a$ [\r{A}] & 3.6470 & 3.5630 & 3.4790 \\
	$P$ [GPa] & 1.90 & 12.82 & 28.57 \\\hline
	$(C_{11}-C_{12})/2$ [GPa] & 18.80 & 23.50 & 29.70 \\
	$C_{44}$ [GPa] & 75.94 & 96.60 & 129.42 \\
	$K=(C_{11}+2C_{12})/3$ [GPa] & 130.05 & 185.05 & 257.31 \\\hline
	$C_{111}-3C_{112}+2C_{123}$ [GPa] & 958 & 1161 & 1454 \\
	$C_{456}$ [GPa] & 47 & 43 & 73\\
	$C_{144}-C_{166}$ [GPa] & 556 & 741 & 957 \\
	$C_{111} + 6C_{112} + 2C_{123}$ [GPa] & $-7393$ & $-9750$ & $-12955$ \\
	$C_{111}$ [GPa] & $-885$ & $-2564$ & $-3826$\\
	$\tfrac13C_{111} + C_{112} + 2(C_{144} + C_{166})$ [GPa] & $-2653$ & $-3759$ & $-5051$
\end{tabular}
\caption{Calculated SOECs and TOECs for Cu at different pressures at 600K.}
\label{tab:results-Cu-600K_P}
}
\end{table}

\begin{table}[h!t!b]
{\renewcommand{\arraystretch}{1.2}
\small
\centering
\begin{tabular}{l|c|c|c}
	$T=600$K & Al & Al & Al\\\hline
	$\rho$ [g/cm$^3$] & 2.668 & 2.965 & 3.660\\
	$a$ [\r{A}] & 4.065 & 3.9247 & 3.6585 \\
	$P$ [GPa] & 1.35 & 10.61 & 46.18 \\\hline
	$(C_{11}-C_{12})/2$ [GPa] & 24.43 & 40.86 & 72.36 \\
	$C_{44}$ [GPa] & 34.20 & 55.08 & 139.63 \\
	$K=(C_{11}+2C_{12})/3$ [GPa] & 68.83 & 109.10 & 242.05 \\\hline
	$C_{111}-3C_{112}+2C_{123}$ [GPa] & 1413 & 2176 & 4245 \\
	$C_{456}$ [GPa] & $-21$ & $-63$ & $-151$\\
	$C_{144}-C_{166}$ [GPa] & $-3$ & 50 & 411 \\
	$C_{111} + 6C_{112} + 2C_{123}$ [GPa] & $-3351$ & $-4945$ & $-8477$ \\
	$C_{111}$ [GPa] & 1510 & $-381$ & $-4206$ \\
	$\tfrac13C_{111} + C_{112} + 2(C_{144} + C_{166})$ [GPa] & $-1382$ & $-2747$ & $-5985$ 
\end{tabular}
\caption{Calculated SOECs and TOECs for Al at different pressures at 600K.}
\label{tab:results-Al-600K_P}
}
\end{table}

\begin{table}[h!t!b]
{\renewcommand{\arraystretch}{1.2}
\small
\centering
\begin{tabular}{l|c|c|c}
	$T=800$K & Al & Al & Al\\\hline
	$\rho$ [g/cm$^3$] & 2.668 & 2.965 & 3.660 \\
	$a$ [\r{A}] & 4.065 & 3.9247 & 3.6585 \\
	$P$ [GPa] & 2.23 & 11.46 & 46.94 \\\hline
	$(C_{11}-C_{12})/2$ [GPa] & 25.24 & 41.13 & 73.43 \\
	$C_{44}$ [GPa] & 34.61 & 55.57& 140.92 \\
	$K=(C_{11}+2C_{12})/3$ [GPa] & 68.34 & 108.85 & 241.41 \\\hline
	$C_{111}-3C_{112}+2C_{123}$ [GPa] & 1607 & 2167 & 3888 \\
	$C_{456}$ [GPa] & $-26$ & $-62$ & $-147$ \\
	$C_{144}-C_{166}$ [GPa] & $-23$ & 49 & 411 \\
	$C_{111} + 6C_{112} + 2C_{123}$ [GPa] & $-3702$ & $-4986$ & $-8881$ \\
	$C_{111}$ [GPa] & 1896 & $-446$ & $-5022$ \\
	$\tfrac13C_{111} + C_{112} + 2(C_{144} + C_{166})$ [GPa] & $-1241$ & $-2723$ & $-6245$
\end{tabular}
\caption{Calculated SOECs and TOECs for Al at different pressures at 800K.}
\label{tab:results-Al-800K_P}
}
\end{table}

\begin{table}[h!t!b]
{\renewcommand{\arraystretch}{1.2}
\small
\centering
\begin{tabular}{l|c|c|c}
	$T=900$K & Cu & Cu & Cu \\\hline
	$\rho$ [g/cm$^3$] & 8.701 & 9.331 & 10.024 \\
	$a$ [\r{A}] & 3.6470 & 3.5630 & 3.4790 \\
	$P$ [GPa] & 3.68 & 14.51 & 30.30 \\\hline
	$(C_{11}-C_{12})/2$ [GPa] & 19.00 & 24.61 & 30.53 \\
	$C_{44}$ [GPa] & 74.75 & 99.55 & 131.47 \\
	$K=(C_{11}+2C_{12})/3$ [GPa] & 129.19 & 184.82 & 257.25 \\\hline
	$C_{111}-3C_{112}+2C_{123}$ [GPa] & 972 & 1171 & 1449 \\
	$C_{456}$ [GPa] & 37 & 38 & 59 \\
	$C_{144}-C_{166}$ [GPa] & 572 & 715 & 953 \\
	$C_{111} + 6C_{112} + 2C_{123}$ [GPa] & $-7724$ & $-9902$ & $-12540$ \\
	$C_{111}$ [GPa] & $-161$ & $-2330$ & $-3912$ \\
	$\tfrac13C_{111} + C_{112} + 2(C_{144} + C_{166})$ [GPa] & $-2416$ & $-3732$ & $-5086$
\end{tabular}
\caption{Calculated SOECs and TOECs for Cu at different pressures at 900K.}
\label{tab:results-900K_P}
}
\end{table}


Additional subtleties arise for the volume non-preserving deformations.
Since both volume and pressure are changed by the deformation, a correction factor $P\mathrm{d}V$ must be taken into account; see e.g. \cite{Vekilov:2016}.
In particular, $\mathrm{d}V=V-V_0$ and $P$ and $V_0$ are the pressure and unit cell volume of the undeformed cell.
In determining $P$, a further difficulty arises:
VASP gives two pressure outputs, 'total' pressure $P_\txt{tot}$ and 'external' pressure $P_\txt{ext}=P_\txt{tot}-P_\txt{corr}$, where the 'ideal gas correction' $P_\txt{corr} = n\kb T/V_0$ is proportional to the temperature\footnote{
In fact, the VASP developers argue in their manual that $P_\txt{corr} = (1-1/N)n\kb T/V_0$ to avoid overestimating the ideal gas correction.}.
Also note that this ideal gas correction (or thermal pressure) corresponds to the kinetic energy $E_\txt{kin}=\tfrac32\kb T(N-1)/(1.602177\times10^{-19})=\tfrac32 X P_\txt{corr}$.

Furthermore, an additional contribution to the energy, the vibrational Helmholtz free energy, 
needs to be added as well \cite{Wang:2010,Shao:2012,Wu:2014}:
\begin{align}
F_\txt{vib} &= \int_0^\infty \left[\frac12\hbar\omega + \kb T \ln\left(1-e^{-\hbar\omega/\kb T}\right)\right]g(\omega)d\omega
\,,
\end{align}
where $g(\omega)$ denotes the phonon density of states and $\omega$ is the phonon frequency.
VASP cannot calculate this contribution, and other authors use various different codes to compute it more accurately.
Here, it will be sufficient to use an estimate based on the isotropic Debye model.
Note that this route will entail higher uncertainties in those elastic constants that are calculated from volume changing deformations.
Within the Debye model, the vibrational Helmholtz free energy can be determined as a function of the Debye temperature $\Theta_D$, namely \cite{Blanco:2004,Shang:2010,Wang:2013}
\begin{align}
 F_\txt{vib} &= \frac98\kb \Theta_D + \kb T\left(3\ln\left[1-e^{-\Theta_D/T}\right] - D\left(\frac{\Theta_D}{T}\right)\right)
\,, \label{eq:FvibDebye}
\end{align}
where
\begin{align}
 D(x) &= \frac{3}{x^3}\int_0^x\frac{t^3}{e^t-1}dt
 \,,
\end{align}
is the Debye function.
The Debye temperature can be calculated from the Debye-Gr{\"u}neisen model \cite{Moruzzi:1988}, though we presently only need to know the change in $F_\txt{vib}$ as a function of volume.
By definition, the change in $\Theta_D$ as a function of volume is related to the Gr{\"u}neisen parameter $\gamma$ as
\begin{align}
\gamma = -\frac{\partial\ln\Theta_D}{\partial\ln V}
\,.
\end{align}
Thus, in the high temperature regime where the first term in \eqref{eq:FvibDebye} can be neglected, we find that
\begin{align}
\frac{\partial F_\txt{vib}}{\partial V}dV &\approx 2\gamma\kb T
\,,
\end{align}
which is proportional to the kinetic energy $E_\txt{kin}$ introduced above.
The energies we computed with VASP hence need to be additionally corrected by
\begin{align}
\frac{4}{3} \gamma \frac{E_\txt{kin}}{X}dV
\,. \label{eq:Pcorr2}
\end{align}
Figure \ref{fig:TOEC1-TOEC6} clearly shows the need for this additional correction for one type of volume changing deformation in copper at ambient pressure and room temperature ('ambient' referring to small $P_\txt{ext}$ prior to applying \eqref{eq:Pcorr2}).
The same correction was also applied to $E_\txt{bulk}$ shown in Figure \ref{fig:bulk}.
Empirically we find $\gamma\approx2.25$ to work well for both copper and aluminum for all three deformation types requiring this correction (see Eqs. \eqref{eq:rescale}, \eqref{eq:TOEC1} , and \eqref{eq:TOEC6}) and at all temperatures and pressures presented here.
In other words, setting $\gamma\approx2.25$ within correction \eqref{eq:Pcorr2} yields energy over deformation data whose lowest energy value coincides with zero deformation for all (volume non-preserving) deformation types and across all temperatures and pressures for Al and Cu considered in this work.
This value for $\gamma$ is indeed close to what we would expect for both Al and Cu:
Indeed, the experimental room temperature (and zero pressure) values for the Gr{\"u}neisen parameter $\gamma$ 
range from 2.00--2.19 for copper and 2.2--2.5 for aluminum, see e.g. Refs. \cite{Gschneidner:1964,Girifalco:1997,Burakovsky:2004}.
Within the very simple Debye model, where $\Theta_D\propto\left( V_0/V \right)^\gamma$ and the Gr{\"u}neisen parameter is strictly constant, we cannot expect perfect agreement and given that the additional energy contribution $F_\txt{vib}$ must lead to energies for the deformed lattice that are always higher than in the undeformed case, it makes sense to treat $\gamma$ as an empirical parameter in the present context even though this approximation leads to higher uncertainties in some of the elastic constants, most notably $C_{111}$.

Additional simulations were carried out for both Cu and Al at several combinations of temperature and pressure.
The results of these simulations are presented in Tables \ref{tab:results-300K}--\ref{tab:results-900K_P}.

\section{Dislocation drag and its sensitivity to temperature/density dependent TOECs}

\begin{figure*}[!h!t]
	\centering
	\includegraphics[width=0.50\textwidth]{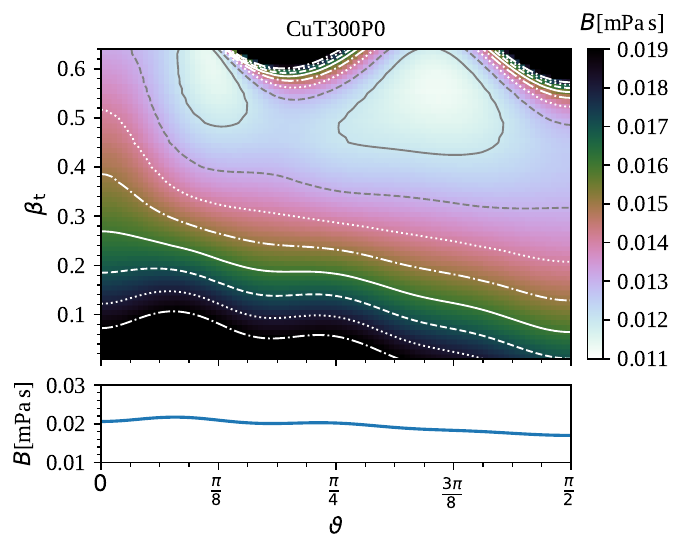}%
	\includegraphics[width=0.50\textwidth]{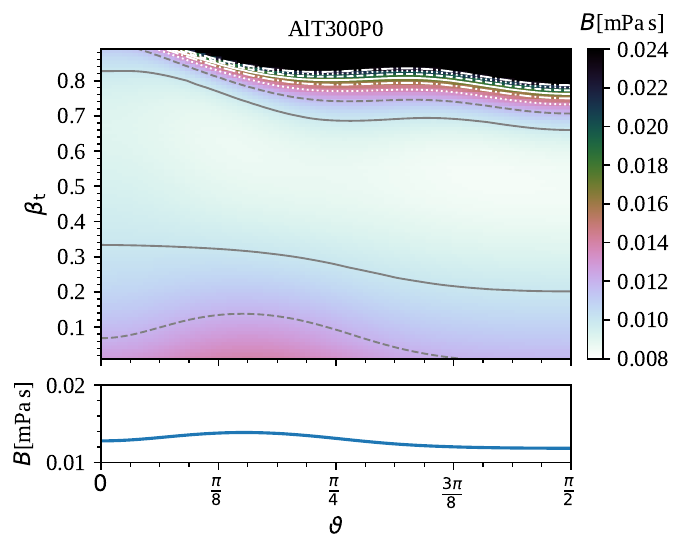}
	\includegraphics[width=0.50\textwidth]{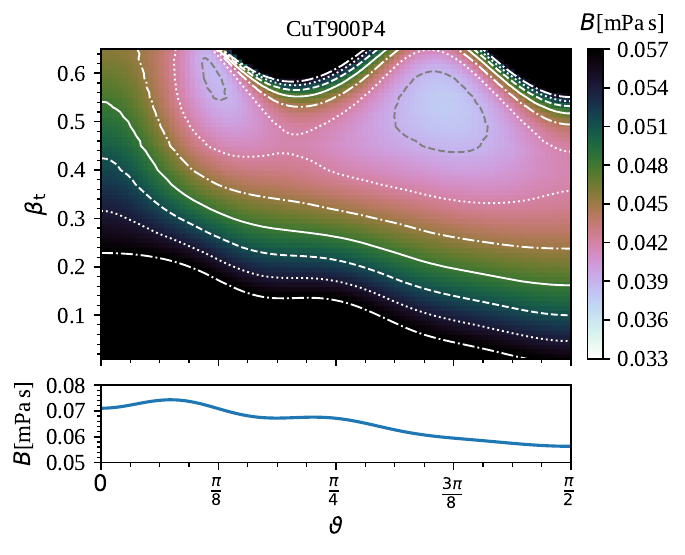}%
	\includegraphics[width=0.50\textwidth]{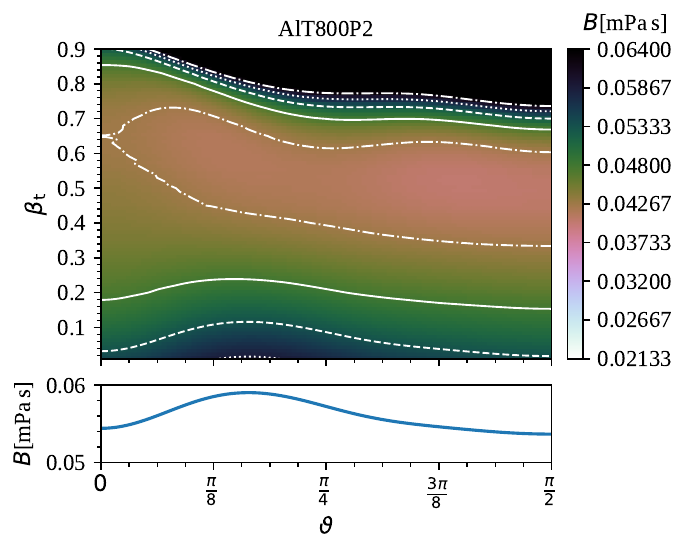}
	\caption{In the top row we show $B(\vartheta,v)$ at ambient conditions for Al and Cu as calculated using the material data determined with VASP as given in Table \ref{tab:results-300K}.
	In the bottom row we show $B(\vartheta,v)$ at high temperature ($T=900$K for Cu, $T=800$K for Al) and the same density as in the top row using the data of Tables \ref{tab:results-Al-800K_P} and \ref{tab:results-900K_P}.
    Since the color scale in the bottom row is changed by $T/300$ compared to the top row, we clearly see that the drag coefficient is slightly enhanced at elevated temperatures relative to linear scaling.
    In each of the four plots, we also show $B(\vartheta)$ at $\bt=v/\ct=0.01$ in the respective lower panels, where velocity $\bt$ was normalized by the polycrystalline average transverse sound speed $\ct$, calculated by averaging over the SOEC using Kr{\"o}ners method \cite{Kroener:1958}; see also \cite{Blaschke:2017Poly}.}
	\label{fig:ambientdrag2d}
\end{figure*}

\begin{figure*}[!h!t]
	\centering
	\includegraphics[width=0.5\textwidth]{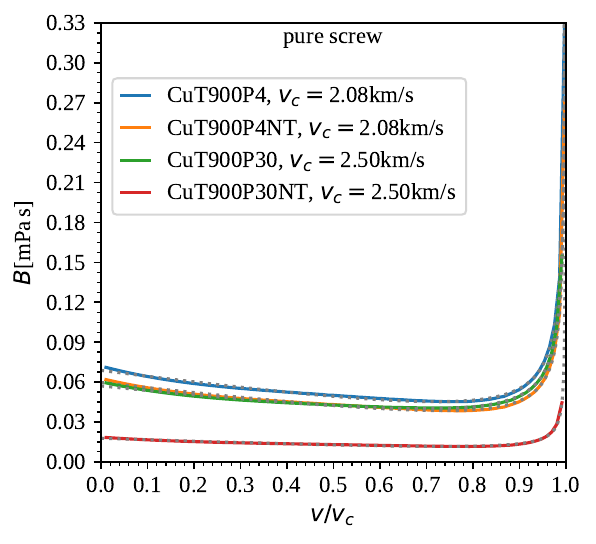}%
	\includegraphics[width=0.5\textwidth]{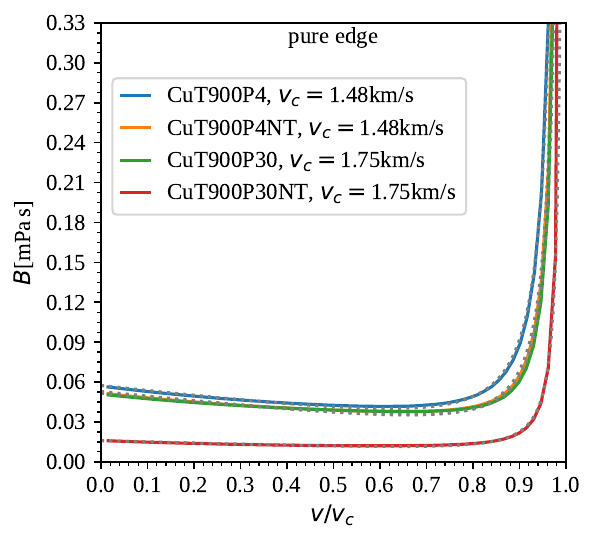}
	\caption{Variation of the dislocation drag coefficient as a function of velocity with density at $T=900$K for pure screw (left) and pure edge (right) dislocations in copper.
	The colored lines were determined numerically using PyDislocDyn whereas the gray dashed lines correspond to fitting functions of the form \eqref{eq:fittingfcts} that were used to numerically derive $B(\sigma)$.
	The labels encode the different pressures (rounded to 1GPa) that were computed and those marked with `NT' were computed with ambient condition TOECs.
	The data used for these calculations are presented in Table \ref{tab:results-900K_P} which also shows which density corresponds to which pressure.
    The dislocation velocity is normalized by the critical velocity.}
	\label{fig:density_drag_900K_fit}
\end{figure*}

\begin{figure*}[!h!t]
\centering
\includegraphics[width=0.5\textwidth]{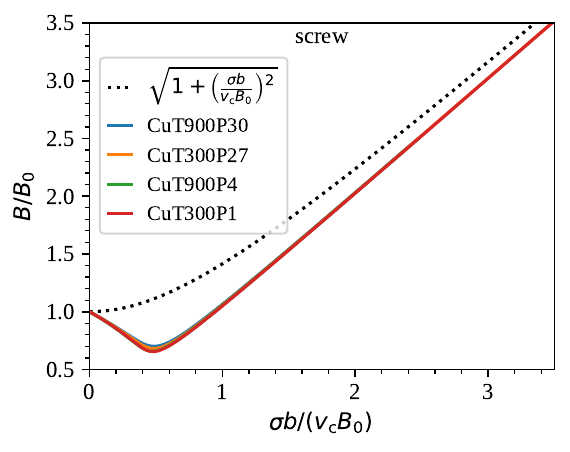}%
\includegraphics[width=0.5\textwidth]{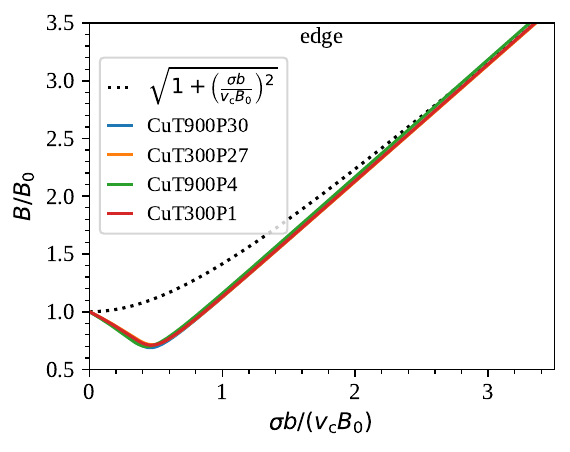}
\caption{We show how accurate the approximation \eqref{eq:Bofsigmasqrt} for the dislocation drag coefficient (black dashed lines) is compared to its numerically determined counter part (colored lines, determined using PyDislocDyn) at the example of Cu and the special cases of pure screw and pure edge dislocations.
The labels encode the different temperatures and pressures (rounded to 1GPa) that were computed and parameters $B_0$ and $v_c$ are listed in Table \ref{tab:dragresults}.}
\label{fig:density_drag}
\end{figure*}

\begin{figure}[!h!t]
\centering
\includegraphics[width=0.5\textwidth]{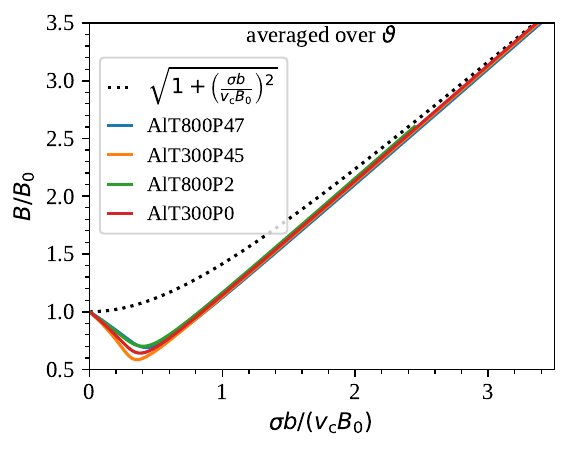}
\caption{We show how accurate the approximation \eqref{eq:Bofsigmasqrt} for the dislocation drag coefficient (black dashed lines) is compared to its numerically determined counter part (colored lines, determined using PyDislocDyn) at the example of Al after averaging over all dislocation character angles.
The labels encode the different temperatures and pressures (rounded to 1GPa) that were computed and parameters $B_0$ and $v_c$ are listed in Table \ref{tab:dragresults}.}
\label{fig:density_dragAl}
\end{figure}

Using the elastic constants determined in the previous section, we now proceed to calculate the dislocation drag coefficient at various points in temperature and density space.
For this purpose we employ the numerical implementation of the phonon wind theory presented in Refs. \cite{Blaschke:2018anis,Blaschke:2019Bpap,Blaschke:2019fits}, namely
the open-source code PyDislocDyn \cite{pydislocdyn}, which was developed by one of us.
The dislocation drag coefficient $B(\vartheta,v)$ was computed for 91 dislocation character angles $\vartheta\in[0,\pi/2]$ and 99 velocities ranging from 1\% to 99\% of the polycrystalline average transverse sound speed $\ct$, which is consistent with the present approximation of an isotropic phonon spectrum.
In Figure \ref{fig:ambientdrag2d} we present as an example, results for $B(\vartheta,v)$ for Cu and Al, each at two different temperatures but the same density (which corresponds to the density they have at low pressure at 300K).

In general, the calculated results for $B(\vartheta,v)$ can be accurately fit by the following functions \cite{Blaschke:2018anis}, as shown e.g. in Figure \ref{fig:density_drag_900K_fit}
\begin{align}
	B(\vartheta,v)&\approx C_0 - C_1\beta +C_2\log\!\left(1-\beta^2\right) +C_3\left({\left(1-\beta^2\right)^{-1/2}}-1\right) +C_4\left({\left(1-\beta^2\right)^{-3/2}}-1\right)
	\,, \label{eq:fittingfcts}
\end{align}
where $\beta=v/v_\txt{c}(\vartheta)$.
As discussed in Refs. \cite{Blaschke:2019a,Blaschke:2021impact}, $B(v)$ diverges at a critical (or limiting) velocity $v_\text{c}$ that is a function of the dislocation character angle $\vartheta$.
This critical velocity is due to a divergence in the dislocation field when the dislocation core is neglected \cite{Markenscoff:2008,Pellegrini:2018,Pellegrini:2020}, thus these critical velocities need not be viewed as hard barriers.
However, the only evidence for the existence of transonic dislocations in metals\footnote{The only experimental evidence for supersonic dislocations was found for a plasma crystal in Ref. \cite{Nosenko:2007}.
No direct measurement of supersonic dislocations in metals exists to the best of our knowledge.}
comes from molecular dynamics simulations \cite{Olmsted:2005,Marian:2006,Daphalapurkar:2014,Tsuzuki:2008,Tsuzuki:2009,Oren:2017,Ruestes:2015,Gumbsch:1999,Li:2002,Jin:2008,Peng:2019,Blaschke:2020MD}.
Analytic expressions for the critical velocities of pure screw ($\vartheta=0$) and pure edge ($\vartheta=\pi/2$) dislocations in fcc slip systems are \cite{Blaschke:2020MD,Blaschke:2021impact}:
\begin{align}
	v_\text{c}^\text{screw} &= \sqrt{\frac{3c'c_{44}}{\rho(c_{44}+2c')}}
	\,,&
    v_\text{c}^\text{edge} &= \sqrt{\frac{\mathrm{min}(c',c_{44})}{\rho}}
    \,,
\end{align}
where $c'=(c_{11}-c_{12})/2$ and $\rho$ is the material density.
The smallest critical velocity vs. $\vartheta$ for fcc slip systems is that for pure edge dislocations \cite{Blaschke:2017lten} and this is therefore also the relevant critical velocity for the drag coefficient averaged over all character angles:
After all, near the smallest critical velocity, the value of $B$ averaged over all character angles will be dominated by its value at the character angle where $B$ diverges.
Note that while $v_\text{c}^\text{edge}$ coincides with the lowest shear wave speed of sound waves traveling in the same direction as a gliding fcc edge dislocation, this is \emph{not} true for $v_\text{c}^\text{screw}$ which is in fact \emph{larger} than the lowest shear wave speed of sound waves traveling in the glide direction of pure fcc screw dislocations; see Ref. \cite{Blaschke:2020MD}.

\begin{table*}[h!t!b]
{\renewcommand{\arraystretch}{1.2}
\small
\centering
\begin{tabular}{l|c|c|c|c|c|c|c}
& P=1GPa & P=3GPa & P=11GPa & P=27GPa & P=4GPa & P=15GPa & P=30GPa \\
Cu & T=300K & T=300K & T=300K & T=300K & T=900K & T=900K & T=900K \\\hline
$v_c^\txt{s}$ [km/s]& 2.08 & 2.11 & 2.26 & 2.47 & 2.08 & 2.30 & 2.50 \\
$B_0^\txt{s}$ [$\mu$Pas] & 19.9 & 21.3 & 19.4 & 17.9 & 68.8 & 54.9 & 57.1 \\
$B_0^\txt{s,NT}$ [$\mu$Pas] & 18.6 & 17.1 & 10.7 & 5.9 & 59.7 & 31.0 & 17.7 \\\hline
$v_c^\txt{e}$ [km/s] & 1.46 & 1.48 & 1.59 & 1.72 & 1.48 & 1.62 & 1.75 \\
$B_0^\txt{e}$ [$\mu$Pas] & 17.1 & 18.8 & 17.1 & 15.6 & 57.4 & 47.7 & 51.7 \\
$B_0^\txt{e,NT}$ [$\mu$Pas] & 16.5 & 14.9 & 9.6 & 5.4 & 52.9 & 26.9 & 15.9 \\\hline\hline
& P=0GPa & P=9GPa & P=45GPa & P=201GPa & P=2GPa & P=11GPa & P=47GPa \\
Al & T=300K & T=300K & T=300K & T=300K & T=800K & T=800K & T=800K \\\hline
$v_c^\txt{av}$ [km/s]& 2.99 & 3.73 & 4.97 & 5.48 & 3.08 & 3.73 & 4.48 \\
$B_0^\txt{av}$ [$\mu$Pas] & 13.1 & 7.0 & 3.4 & 3.3 & 57.3 & 18.8 & 17.9 \\
$B_0^\txt{av,NT}$ [$\mu$Pas] & 13.1 & 5.2 & 1.3 & 0.8 & 36.1 & 13.9 & 4.4
\end{tabular}
\caption{Drag coefficient at low velocity and limiting dislocation velocity for Cu and Al at different temperatures and pressures (rounded to 1GPa).
Values marked with `NT' were computed with ambient condition TOECs.
Superscripts `s',  `e', and `av' refer to `screw', `edge', and `averaged over all character angles'.
The data used for these calculations are presented in Tables \ref{tab:results-Cu-300K_highP}, \ref{tab:results-Al-300K_highP}, \ref{tab:results-Al-800K_P}, and \ref{tab:results-900K_P}, which also show which density corresponds to which pressure.}
\label{tab:dragresults}
}
\end{table*}

Using the defining equation for dislocation drag, $b\sigma = v B$, where $\sigma$ denotes the resolved shear stress, we numerically solve for the dislocation velocity as a function of the stress hence the drag coefficient as a function of the stress (see \cite{Blaschke:2019a,Blaschke:2021impact}).
In principle, this can be done for any dislocation character, but here, in Figure \ref{fig:density_drag} and Table \ref{tab:dragresults}, we present $B(\sigma)$ only for pure screw and edge dislocations in copper. 
Additionally, we show the average of the drag coefficient over all character angles as a function of the stress for Al in Figure \ref{fig:density_dragAl} and Table \ref{tab:dragresults}. 

In Refs. \cite{Blaschke:2019a,Blaschke:2021impact} we showed that the drag coefficient in the asymptotic regime grows linearly with resolved shear stress and its slope is equal to the ratio $b/v_\text{c}$ of Burgers vector magnitude to critical velocity.
Consequently, we expect that the drag coefficient is sensitive to the TOEC only at low and intermediate velocities, but not at very high stresses.
Since plastic deformation is very sensitive to $B$ only at high stresses \cite{Blaschke:2019a}, we argue that the TOEC need only be determined at one point in $(T,\rho)$ space (TOEC data are typically available only for ambient conditions), without introducing a significant error when computing strain rates using $B$ determined from TOEC at a different point in temperature-density space.
It was argued in Refs. \cite{Blaschke:2019a,Blaschke:2021impact} that the following approximation for $B$ is sufficient for many applications
\begin{align}
	B(\vartheta,\sigma) &= B_0\sqrt{1+\left(\frac{b\sigma}{v_\text{c} B_0}\right)^2}
	\,, \label{eq:Bofsigmasqrt}
\end{align}
where $B_0$ is an appropriate value in the interval $[B(v=0),\text{min}\left(B(v)\right)]$; see \cite{Blaschke:2019a}.
The computationally least expensive choice is $B_0=B(\vartheta,v=0)$, which we adopt here since it was argued to be sufficiently accurate for high strain rate applications \cite{Blaschke:2021impact}.
The function $B_0(\vartheta)$ must be calculated from the full drag coefficient theory, whereas $v_\text{c}(\vartheta)$ can be determined for any character angle and slip system by solving an eigenvalue problem of a (character dependent) $3\times3$ matrix as discussed in \cite{Blaschke:2017lten}.
The solution for pure screw and edge dislocations in fcc systems can be derived analytically and is given by \eqnref{eq:Bofsigmasqrt}.
All SOECs and the material density for the temperature of interest are needed for accuracy.
Only $B_0$ requires knowledge of the TOECs; the critical velocity $v_c(\vartheta)$ does not depend on the TOEC, as elucidated earlier.  
Since the high-rate deformation is insensitive to $B_0$, it is usually sufficient to calculate $B_0$ using the ambient TOECs.
To support this view, we present as an example Figure \ref{fig:density_drag_900K_fit} where we compare the drag coefficient computed with the actual TOECs at 900K and two different pressures (determined from VASP) to the drag coefficient computed using the ambient condition TOECs (but the SOECs for the temperatures and densities shown).
As the velocity $v$ approaches the critical velocity $v_c$, the drag coefficient $B$ tends to infinity and the finite difference between the values for $B(v\to v_c)$ calculated with the two different sets of TOEC is much less than either value of $B$.
As the pressure increases to values $\gtrsim10$GPa, this finite difference between the values for $B(v\to v_c)$ calculated with the two different sets of TOEC increases, and thus the full set of TOECs should be computed for pressures sufficiently close ($\approx\pm10$GPa) to the pressure of interest for improved accuracy.

The approximation \eqref{eq:Bofsigmasqrt} is compared to the more accurate numerical results in Figures \ref{fig:density_drag} and  \ref{fig:density_dragAl} .
We see that for a range of temperatures and pressures the curves almost retain their shape upon normalizating drag coefficient $B$ by $B_0$ and stress $\sigma$ by $v_cB_0/b$.

\subsubsection*{Comparison to experimental and simulation data for dislocation drag:}

Direct comparison of drag coefficient $B$ to experiments is limited to the `viscous' regime of $\bt\sim0.01$ since experimental data at higher dislocation velocities are not available so far.
Also, the available experimental data cannot resolve dislocation character dependence and to our knowledge, no experiments have been carried out at temperatures higher than room temperature or high pressure.
As was already pointed out in Refs.~\cite{Blaschke:2018anis,Blaschke:2019fits,Blaschke:2019Bpap}, our predictions for $B(\bt\sim0.01)$ at room temperature and low pressure agree well with experimental results for both Al and Cu.
In particular, the experimental data  for Al range from $\sim0.005\,$mPas to $\sim0.06\,$mPas, cf.~\cite{Hikata:1970,Gorman:1969,Parameswaran:1972} and those for Cu range from $\sim0.0079\,$mPas to $\sim0.08\,$mPas, cf.~\cite{Suzuki:1964,Zaretsky:2013,Stern:1962,Greenman:1967,Alers:1961}.
Our theoretical predictions (as a function of dislocation character) are shown in the lower panel of the top row of Figure \ref{fig:ambientdrag2d}.
Molecular dynamics (MD) predictions for $B$ in the viscous regime are also available in the literature.

MD simulation results at room temperature and low pressure are in the range $\sim0.007$--$0.2\,$mPas for Al~\cite{Olmsted:2005,Yanilkin:2014,Cho:2017}, and $\sim0.016$--$0.022\,$mPas for Cu~\cite{Oren:2017,Wang:2008}.
It has also been shown within MD simulation studies \cite{Yanilkin:2014,Krasnikov:2010} that the temperature dependence of $B$ at elevated temperatures $T>300$K is roughly linear to a first order approximation, though slightly enhanced at large $T$ --- a finding we confirm within our theoretical calculations.
We are not aware of any systematic (experimental or MD simulation) study of $B$ as a function of pressure.
As for the high dislocation velocity regime, MD simulations \cite{Oren:2017} have found that edge dislocations in Cu asymptotically approach the theoretically predicted limiting velocity with increasing stress, which implies a steep rise in drag coefficient $B$ in this regime (consistent with our present predictions).
On the other hand, those simulations show that above some very high critical stress ($\sigma\gg1$GPa, i.e. $\sigma b/(v_cB_0)\gg 10$ which is much higher than the range shown in Fig. \ref{fig:density_drag}), edge dislocations can also jump to ``supersonic'' speeds, see \cite{Oren:2017}.
Such supersonic speeds have been found for screw dislocations in Cu only at very low temperatures, but not at room temperature (despite earlier claims): this has been recently clarified in Ref. \cite{Blaschke:2020MD}.
In Al, on the other hand, various instabilities (such as dislocation nucleation) prevent dislocations of any character from really reaching the limiting velocity within MD simulations \cite{Vandersall:2004,Olmsted:2005,Blaschke:2020MD}.
Furthermore, no experimental evidence has been found that supersonic dislocations truly occur in metals.
Though certainly interesting, the question of under which conditions supersonic dislocation motion may be possible, requires further theoretical investigation which is beyond the scope of this work.

\section{Conclusion}

A first-principles calculation of dislocation drag from phonon wind requires, as input parameters, the material density as well as second- and third-order elastic constants.
Experimental data on the latter are scarce and typically only available for ambient conditions.
Determination of the third-order elastic constants at different temperatures and densities is computationally very expensive.
In this paper, we have computed SOEC and TOEC and subsequently the anisotropic dislocation-character and stress-dependent drag coefficients at several points in temperature and density space for aluminum and copper.
Following a sensitivity study, we subsequently argued that small differences in the TOEC are unimportant in the very high strain rate regime.
Together with a simple analytical approximation for dislocation drag as a function of character angle and stress, sufficiently accurate results can be achieved at significantly lower computational cost,
since the TOECs need only be computed sparsely in temperature and pressure space.
Finally, we also note that our computations show that the drag coefficient changes non-linearly with pressure and temperature.
In fact, the temperature dependence $B(T)$ is enhanced compared to the first-order approximation of linear $T$-dependence.
However, its (non-linear) pressure dependence, $B(P$), does not seem to solve the open question as to why the yield strength $Y$ scales non-linearly with the shear modulus $G$ for Cu, as discussed in the introduction.
As for the other two cases where experiments found additional hardening, namely, Ta and Pb, a more detailed study of phonon drag may be required, although our preliminary results using the elastic constants of Ref. \cite{Krasilnikov:2012} show that in the case of Ta, too, phonon drag does not seem to the leading hardening mechanism.
In fact, stiffer elastic moduli lead not only to higher sound speeds but also to higher critical dislocation velocities and thus to a lower drag coefficient in the high strain rate regime.

\subsection*{Acknowledgments}
\noindent
DNB would like to thank Ann E. Mattsson for related discussions on VASP.
This work was performed under the auspices of the U.S. Department of Energy under contract 89233218CNA000001.
In particular, the authors are grateful for the support of the Physics and Engineering Models sub-program element of the Advanced Simulation and Computing program.

\bibliographystyle{utphys-custom}
\bibliography{dislocations}
\end{document}